\documentclass[a4paper]{jpconf}
\usepackage{graphicx,amssymb,wrapfig}
\begin{document}

\def\cO#1{{\cal{O}}\left(#1\right)}
\newcommand{\cp}{\hspace{1mm}\mbox{\raisebox{.3mm}{$\diagup$} \hspace{-7.3mm} C \hspace{-3.2mm} P}\hspace{1mm}}
\newcommand{\bc}{\begin{center}}
\newcommand{\ec}{\end{center}}
\newcommand{\ul}{\underline}
\newcommand{\ol}{\overline}
\newcommand{\ra}{\rightarrow}
\newcommand{\sm}{SM~}
\newcommand{\as}{\alpha_s}
\newcommand{\aem}{\alpha_{em}}
% maggie added
\newcommand{\ta}{\theta_1}
\newcommand{\tb}{\theta_2}
\newcommand{\ga}{\gamma_a}
\newcommand{\gb}{\gamma_b}
\newcommand{\bea}{\beta_1}
\newcommand{\beb}{\beta_2}
\newcommand{\cph}{c_\phi}

\catcode`\@=11
% ----------------------------------------------------------------------------
% Original Latex definition of citex, except for the removal of
% 'space' following a ','

\def\@citex[#1]#2{\if@filesw\immediate\write\@auxout{\string\citation{#2}}\fi
  \def\@citea{}\@cite{\@for\@citeb:=#2\do
    {\@citea\def\@citea{,\penalty\@m}\@ifundefined
       {b@\@citeb}{{\bf ?}\@warning
       {Citation `\@citeb' on page \thepage \space undefined}}%
\hbox{\csname b@\@citeb\endcsname}}}{#1}}

\def\citer{\@ifnextchar [{\@tempswatrue\@citexr}{\@tempswafalse\@citexr[]}}
% ----------------------------------------------------------------------------
% \citer as abbreviation for 'citerange' replaces the ',' by a '--'
%

\def\@citexr[#1]#2{\if@filesw\immediate\write\@auxout{\string\citation{#2}}\fi
  \def\@citea{}\@cite{\@for\@citeb:=#2\do
    {\@citea\def\@citea{--\penalty\@m}\@ifundefined
       {b@\@citeb}{{\bf ?}\@warning
       {Citation `\@citeb' on page \thepage \space undefined}}%
\hbox{\csname b@\@citeb\endcsname}}}{#1}}
% ----------------------------------------------------------------------------
\catcode`\@=12
% end maggie added
\def\Ecm{\ifmmode{E_{\mathrm{cm}}}\else{$E_{\mathrm{cm}}$}\fi}
\def\lsim{\buildrel{\scriptscriptstyle <}\over{\scriptscriptstyle\sim}}
\def\gsim{\buildrel{\scriptscriptstyle >}\over{\scriptscriptstyle\sim}}
\def \lum{{\cal L}}

\def\lapp{\mathrel{\rlap{\raise.5ex\hbox{$<$}}
                    {\lower.5ex\hbox{$\sim$}}}}
\def\gapp{\mathrel{\rlap{\raise.5ex\hbox{$>$}}
                    {\lower.5ex\hbox{$\sim$}}}}
\newcommand{\decay}[2]{
\begin{picture}(25,20)(-3,3)
\put(0,-20){\line(0,1){10}}
\put(0,-20){\vector(1,0){15}}
\put(0,0){\makebox(0,0)[lb]{\ensuremath{#1}}}
\put(25,-20){\makebox(0,0)[lc]{\ensuremath{#2}}}
\end{picture}}

\title{CP Asymmetries in Higgs decays to ZZ at the LHC}

\author{Rohini M. Godbole$^1$, David J. Miller$^2$, M.~Margarete~M\"uhlleitner$^{3,4}$}

\address{$^1$ Centre for High Energy Physics, Indian Institute of Science, Bangalore, 560 012, India.}
\address{$^2$ Dept. of Physics and Astronomy, University of Glasgow, Glasgow G12 8QQ, U.K.}
\address{$^3$ Theory Division, Physics Department, CERN, CH-1211 Geneva 23, Switzerland.}
\address{$^4$ Laboratoire d'Annecy-Le-Vieux de Physique Th\'eorique, LAPTH, France.}

\begin{abstract}
We examine the effect of a general HZZ coupling through a study of the
Higgs decay to leptons via Z bosons at the LHC.  We discuss various
methods for placing limits on additional couplings, including
measurement of the partial width, threshold scans, and asymmetries
constructed from angular observables. We find that only the asymmetries
provide a definitive test of additional couplings. We further estimate the
significances they provide.
\end{abstract}

\section{Introduction}
The verification of the Higgs mechanism as the cause of electroweak
symmetry breaking and the discovery of the Higgs boson is the next big
goal of particle physics. However, it is not enough to simply find a
new resonance in the Higgs search channels at the next generation of
colliders. One must ensure that this resonance is indeed the Higgs
boson by measuring its properties: its CP and spin, to demonstrate its
predicted scalar nature; its couplings to known particles, to verify
that these couplings are proportional to the particle's mass; and the
Higgs self couplings, in order to reconstruct the Higgs potential
itself. This will be a challenging programme and will not
be fully realised at the Large Hadron Collider (LHC) (e.g.\ the
quartic Higgs self coupling will be out of reach). However,
such an analysis will be crucial in our investigation of electroweak
symmetry breaking in scenarios where the suspected Higgs boson is all
we find at the LHC, as well as scenarios where new physics is
discovered. In the former case, testing for deviations from the
Standard Model (SM) may provide clues to resolving some of the SM's
long standing problems; in the latter case, the Higgs boson properties
will provide essential information on the nature of the new physics.

It is interesting to note that the Higgs boson's CP (and spin) is
intimately related to its couplings to other SM particles, since its
scalar or pseudoscalar nature allow or forbid certain tensor
structures in the Higgs boson couplings. In this report, we investigate
the tensor structure of the $HZZ$ vertex in order to shed some light
on the Higgs boson's CP. We write down the most general tensor vertex
for this coupling and investigate how the additional terms influence
the decay $H \to ZZ^{(*)} \to 4$~{\it leptons} at the LHC. For a more
detailed description of this analysis, see Ref.\cite{Godbole:2007cn}.

The most general vertex for a spinless particle coupling to a pair of
$Z$ bosons, with four-momenta $q_1$ and $q_2$, is given by,
\begin{eqnarray} V_{HZZ}^{\mu \nu} \, =\,
\frac{ig m_Z}{\cos\theta_W} \left[ \,a\, g_{\mu\nu} 
+  b \,\frac{p_\mu p_\nu}{m_Z^2}  
+  c \,\epsilon_{\mu\nu\alpha\beta} \, \frac{ p^\alpha k^\beta}{m_Z^2} 
\, \right],
\label{vertex}
\end{eqnarray}
where $p=q_1+q_2$ and $k=q_1-q_2$, $\theta_W$ denotes the weak-mixing
angle and $\epsilon_{\mu \nu\alpha\beta}$ is the totally antisymmetric
tensor with $\epsilon_{0123}=1$. The CP conserving tree-level Standard
Model coupling is recovered for $a=1$ and $b=c=0$.

Terms containing $a$ and $b$ are associated with the coupling of a
CP-even Higgs, while that containing $c$ is associated with that of a
CP-odd Higgs boson. The simulanteous appearance of a non-zero $a$
(and/or $b$) together with a non-zero $c$ would lead to CP
violation. In general these parameters can be momentum-dependent form
factors that may be generated from loops containing new heavy
particles or equivalently from the integration over heavy degrees of
freedom giving rise to higher dimensional operators.  The form factors
$b$ and $c$ may, in general, be complex, but since an overall phase
will not affect the observables studied here, we are free to adopt the
convention that $a$ is real.

\section{The total width}
\begin{figure}[htb]
\begin{minipage}{20pc}
\includegraphics[width=20pc,clip=true]{thresh.eps}
\caption{\label{thresh}The dependence of the Higgs decay width on the
virtuality of the most virtual $Z$ boson.}
\end{minipage}\hspace*{3pc}%
\begin{minipage}{13pc}
\includegraphics[width=13pc,clip=true]{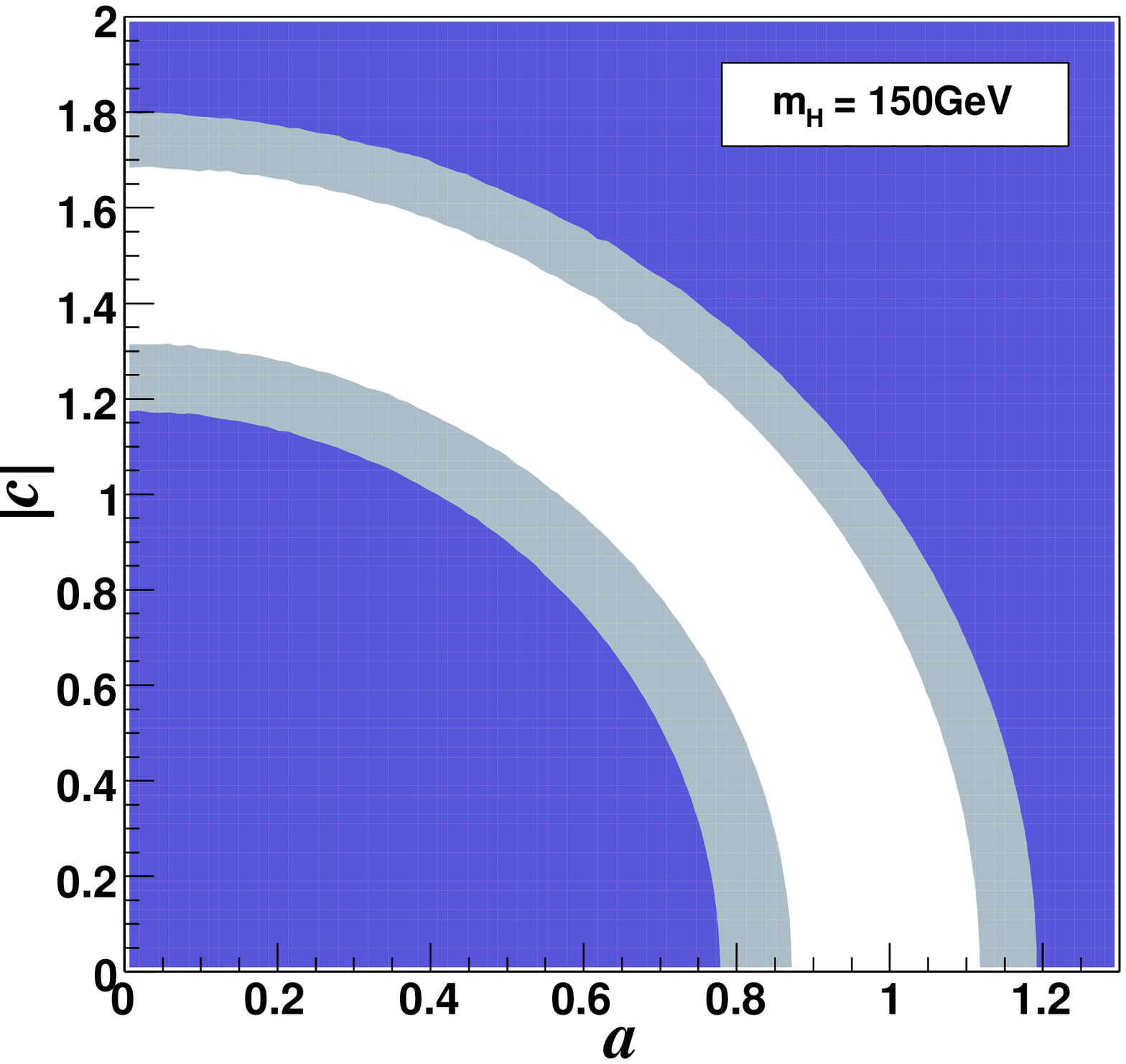} \vspace*{-2mm}
\caption{\label{scan}The deviation of the width from the SM prediction.}
\end{minipage} 
\end{figure}

One method of investigating the tensor structure of the $HZZ$ coupling
is to examine the threshold behaviour of the decay $H \to
ZZ^*$~\cite{Choi:2002jk}.  Notice that the additional terms in the
vertex all have a momentum dependence and will vanish at
threshold. The SM term does not, and although the SM width will still
vanish at threshold due to a shrinking phase space, it will have a
much steeper slope than a pure CP-odd state.
This can be seen in Figure~\ref{thresh} which shows the dependence of
the partial width of a $150\,$GeV Higgs boson on the virtuality of the
most virtual $Z$ boson. Notice that the pure CP-even (SM) (solid black
curve) and pure CP-odd (dashed blue curve) states are easily
distinguishable. However, when one has a CP-violating combination of
couplings (dot-dashed red curve) the SM terms will be dominant near
threshold and it is very difficult to distinguish from the SM.

Alternatively, one could examine the magnitude of the total decay
width for \mbox{$H \to ZZ^* \to 4$~{\it leptons}} to see if it differs
from the SM. For the vertex of Equation~\ref{vertex}, the dependence
on the coefficients $a$, $b$ and $c$ is given by,
\begin{equation}
\frac{\partial^2 \Gamma_{H}}{\partial q_1^2  \partial q_2^2} \sim
\beta  
\left\{ a^2 \left[\beta^2+\frac{12q_1^2 q_2^2}{m_H^4} 
\right] + 
|b|^2 \frac{m_H^4}{m_Z^4} \frac{\beta^4}{4} 
+ |c|^2 \frac{8 q_1^2 q_2^2}{m_Z^4} \beta^2  +
a\Re e(b) \frac{m_H^2}{m_Z^2} \beta^2  
\sqrt{\beta^2+  \frac{ 4 q_1^2 q_2^2}{m_H^4}} 
\right\},
\label{hwidth}
\end{equation}
where $\beta$ is the usual Lorentz boost factor for the
$Z$-bosons. (Notice that the only term with a linear $\beta$
dependence (from the phase space) is proportional to $a^2$,
illustrating the principle described above for the threshold scan.) If
additional terms are present one expects them to increase or decrease
the width according to this equation. We used the ATLAS study of
Ref.~\cite{tdratl,Hohl:2001atl} (including cuts and efficiencies) to
estimate the number of signal and background events for the SM and
CP-violating scenarios (scaling the signal according to
Equation~\ref{hwidth}). In Figure~\ref{scan} we plot the number of
standard deviations from the SM that the CP-violating scenario would
imply, for a $150\,$GeV Higgs boson and an integrated luminosity of
$300\,{\rm fb}^{-1}$ (we set $b=0$ for simplicity). The white area
represents scenarios where the significance of the deviation is less
than $3\, \sigma$, the light blue/grey region represents a
$3-5\,\sigma$ deviation, while the dark blue/grey region represents a
greater than $5\, \sigma$ deviation. This measurement would allow one
to rule out much of the $a-|c|$ parameter space, but does not allow
one to definitively rule out (or place significant limits on) the
CP-odd coupling $|c|$. A SM-like rate is perfectly consistent with a
large value of $|c|$ and a small value of $a$.

\section{Asymmetries as a probe of CP violation}
To definitively ascertain whether or not extra tensor structures are
present in the $HZZ$ vertex one is better served by measuring
asymmetries which vanish when such terms are absent. Such an asymmetry
can be constructed from an observable, ${\cal O}$, based on the angles
of the final state leptons,
\begin{equation}
{\cal A} = \frac{\Gamma ({\cal O} > 0)-\Gamma ({\cal O}<0)}
{\Gamma ({\cal O} > 0)+\Gamma ({\cal O}<0)} \;.
\label{asymdef}
\end{equation}
The choice ${\cal O}=\cos \theta_1$, where $\theta_i$ is the angle
between lepton $i$ and its parent $Z$'s direction of travel, as
measured in the $Z$ rest frame, provides an asymmetry proportional to
$a\,\Im m(c)$. Unfortunately, for this particular choice the asymmetry
is rather small (always $\lesssim 7\%$), making it a rather poor
discriminant. However, other observables can be constructed which do
much better. For example, ${\cal O}_5 = \sin\theta_1 \sin\theta_2
\sin\phi [\sin\theta_1 \sin\theta_2\cos\phi + \cos\theta_1
\cos\theta_2 ]$, 
where $\phi$ is the azimuthal angle between the two
planes formed by lepton-antilepton pairs,
\begin{wrapfigure}{r}{18pc}
\includegraphics[width=17pc,clip=true]{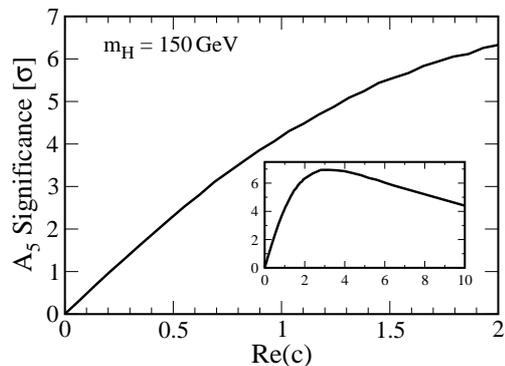}
\caption{\label{a5_sig}The significance of the deviation from zero of
the asymmetry ${\cal A}_5$.}
\end{wrapfigure}
provides a good test of non-zero $\Re e(c)$, with an asymmetry, ${\cal
A}_5$, sometimes as large at $15\%$. We calculated the significance
with which non-zero $\Re e(c)$ would manifest at the LHC by taking the
number of signal and background events from the ATLAS study. Since the
contamination of the asymmetry from the background is rather minimal
we use the number of events after the initial selection cuts, but
before applying additional isolation and impact parameter cuts to
remove the irreducible backgrounds. The significance, for a $150\,$GeV
Higgs boson and a total luminosity of $300\,{\rm fb}^{-1}$, is shown
in Figure~\ref{a5_sig}, and we see this asymmetry would provide a
$3(5) \,\sigma$ exclusion limit for $\Re e(c) \gtrsim 0.66(1.28)$. See
Ref.\cite{Godbole:2007cn} for further details, including additional
asymmetries and an analysis for $m_H=200$GeV.

\end{document}